\documentstyle{elsart}

\begin{document}
\begin{frontmatter}
\title{Explosive Nucleosynthesis: Prospects}
\author[ad1]{David Arnett\thanksref{corr}},
\thanks[corr]{E-mail: darnett@as.arizona.edu}
\address[ad1]{Steward Observatory, University of Arizona,
	 \cty Tucson AZ 85721, \cny USA}

\begin{abstract}

Explosive nucleosynthesis is a combination of the nuclear physics of
thermonuclear reactions, and the hydrodynamics of the plasma in which
the reactions occur. It depends upon the initial conditions---the stellar
evolution up to the explosive instability, and the nature of the explosion
mechanism.

Some key issues for explosive nucleosynthesis are the
interaction of burning with hydrodynamics, the degree
of microscopic mixing in convective zones,
 and the breaking of spherical symmetry by convection and rotation. 
Recent experiments on high intensity lasers provides new
opportunities for laboratory testing of astrophysical hydrodynamic
codes.  Implications of
supernovae 1987A and 1998bw (GRB980425?), and $\eta$~Carina
are discussed, as well as the formation of black holes or
neutron stars.  

\end{abstract}

\begin{keyword}
Nucleosynthesis \sep Hydrodynamics \sep Supernovae \sep Lasers \sep Black Holes
\sep GRB's \sep Rotation \sep Convection \sep Jets
\end{keyword}
\end{frontmatter}

\section{Introduction}

This paper is presented to honor David Norman Schramm. It will be a 
personal view of where this field is going, not a review of work
already done, and in a sense it is a formal
continuation of many previous discussions Dave and I had on this topic.

There is a growing consensus that the
synthesis of the elements and their isotopes may be divided into
three major components: 
\begin{itemize}
\item Cosmological synthesis of the light elements,
\item Hydrostatic synthesis in stars, and
\item Hydrodynamic synthesis in stellar explosions.
\end{itemize}
The last item, ``explosive nucleosynthesis,'' will be the focus here.
Although thermonuclear burning has a morphological equivalent
in terrestrial burning by chemical reactions, this discussion will
be couched in the language of physics, and named dimensionless numbers,
whose meaning may not be obvious to scientists outside the combustion
and/or the fluid dynamics
community, will be used sparingly. These numbers can be constructed
from ratios of the time scales and of length scales discussed below.

Any discussion of nucleosynthesis requires consideration of the empirical
basis of the nuclear reaction rates. The historical context and references
are collected in \cite{da96}. The recent appearance of an excellent review
\cite{ktw98} allows us the freedom here to concentrate instead upon topics
less discussed but of not less importance: the hydrodynamic context of
the burning. However, the brevity of this mention should not be interpreted
as an indication that the construction of an empirical basis for 
nucleosynthesis theory is a finished topic.
To use nucleosynthesis products as a
probe of stellar environments, it is necessary to insure that abundance
features observed and simulated are due to the history of physical 
conditions, not poorly known reaction rates.
Nuclear physics is the foundation upon which we build. 

Explosive nucleosynthesis is a combination of the nuclear physics of
thermonuclear reactions, and the hydrodynamics of the plasma in which
the reactions occur. It depends upon the initial conditions---the stellar
evolution up to the explosive instability, and the nature of the explosion
mechanism.

\section{Relevant Scales of Length and Time}

There is a vast difference between macroscopic and microscopic 
lengths in stars (see \cite{da96}, Chapter 11). 
Burning is a nuclear process which occurs in
a context of enormous dimension.
The radius of the a typical star, the Sun, is $7 \times 10^{10}\rm cm$,
while the internuclear spacing is roughly
$1.2 \times 10^{-8}{\rm\ cm} (A/\rho)^{1/3}$, where $\rho$ is the density
in cgs units and $A$ the mass number of the most abundant nucleus.
The mean density of the Sun is 1.4 g/cc (but in a presupernova
densities increase to above $10^{9}$ g/cc). Suppose that collision
cross sections have a scale of order $10^{-16}\rm\ cm^2$, corresponding
to a collisional mean free path of 
\begin{equation}
\lambda_{col} \approx 1.6 \times 10^{-8} A/\rho \rm\ cm. 
\end{equation}
A characteristic
scale for electron-photon interactions is the Thomson cross section, 
corresponding to a mean free path of roughly 
\begin{equation}
\lambda_{rad} \approx 1.6 A/\rho\rm \ cm.
\end{equation}
The precise determination of cross sections, properly averaged over
the relevant distributions is complex, but the qualitative result is the
same as given by these simple estimates. 
This suggests that thermal energy moves much more easily than composition,
and that our terrestrial intuition with flames may be colored by the ease with
which heat moves on the length scales with which we are familiar.

Consider a homogeneous sphere of radius $R$; its diffusion time is
\begin{equation}
\tau_{dif} = {3 \over \pi^2} R^2/\lambda v = {3 \over \pi^2} R^2 \rho N_A 
\sigma/ v,
\end{equation}
where
 $v$ is the velocity of the diffusing entities, $\sigma$ is the cross section,
and $N_A$ is Avagadro's number. Diffusion is slow for large objects like 
stars. However, time scales for binary interactions are
\begin{equation}
\tau_{reac} \approx 1/(\rho N_A \sigma v). 
\end{equation}
This dramatically different density dependence
makes reactions faster and diffusion slower at higher densities, that is,
at advanced burning stages.

Simulations of stellar hydrodynamics and evolution presume that the stellar
plasma is homogeneous on scales below the size of the computational zone.
However, this is unlikely for the late stages of stellar life, which occur
on time scales of seconds, not billions of years. This scaling with density
insures that the problem of incomplete mixing gets worse as the star
evolves. It is exacerbated in massive stars because they evolve most
rapidly; they are primary sites of explosive nucleosynthesis.

\section{Types of Nuclear Burning in Stars}

The simplest sort of stellar burning is that of a ``radiative zone,'' in
which heat can diffuse in and out, but composition is only changed by
the conversion of fuel to ashes. This is thought to be the case in the
center of the sun, for example, and the driving change which causes such
stars to become red giants.

Even quasistatic 
nuclear burning can give locally intense heating, so that steep temperature
gradients are formed which may drive convection. Convective currents are
likely to be turbulent in stars because their large size and small viscosity
give a large Reynolds number, that is, chaotic flow. This turbulent mixing
is more effective than diffusion. As we shall see, the simulation of this
 represents a major
challenge. Although such burning often is not explosive, it sets the stage
for the explosive events, and seems to be their direct predecessor.

A violent but relatively simple process is detonation \cite{llfm}. The process
is explosive, and supersonic. It
is local in the sense that shock compression heats the fuel to the flash
point, so that it incinerates. The motion of the shock is purely hydrodynamic
except that the shock is enhanced by the energy released by the burning.
The difficulty lies in determining how the detonation begins; this is an
old and complex problem \cite{fd79}.

Another violent type of burning is deflagration \cite{llfm}, in which
new fuel is ignited by heat flow from regions already burned. Unlike
detonation, this depends upon the nature of the heat flow (conduction or
radiative diffusion). Deflagration is subsonic, and in that sense is
milder than detonation, into which it may develop.
Deflagration is often unstable in stars (see Fig.~11.1 in \cite{da96}),
which adds to the complexity of an already complex situation. 

These themes, sufficiently complicated in their own right, are the basis
for explosive nucleosynthesis. This level of complexity, and the need
for quantitative predictions, makes computer simulations a necessary tool.

\section{Laboratory Supernova}

Before trusting hydrodynamic simulations, detailed quantitative testing
of the computational framework is necessary. 
High intensity lasers have been successfully used to study 
the Rayleigh-Taylor (RT)
and Richtmeyer-Meshkov (RM) instabilities well into the nonlinear 
regime \cite{br94}. The RT instability occurs when gravity
tries to pull a heavier fluid through an underlying lighter one. The
RM instability is similar except that the role of gravity is replaced
by the inertial acceleration from the passage of a shock wave.
Core collapse supernovae are driven by a poweful shock,
and such shocks are the breeding ground of hydrodynamic instabilities.
Observations of SN 1987A strongly suggested the occurrence of mixing
of radioactive material outward, a phenomena not seen in one dimensional (1D)
simulations but predicted in 2D simulations \cite{afm89}.

It is prudent to compare astrophysical codes to those
mature computer codes used by the inertial confinement fusion (ICF) 
community  on
problems for which both should be applicable. This has been successfully
done \cite{jk97}; both CALE (ICF) and PROMETHEUS (astrophysics) codes
were used to simulate an experiment on the NOVA laser at Lawrence Livermore
National Laboratory (LLNL). The observed instabilities (RT and RM) were
well simulated by both codes; the bubble and spike positions were reproduced,
even into the strongly nonlinear regime. 

A theoretical look at the relation between the hydrodynamics occurring in
SN~1987A and in the laser experiments shows that a rigorous mapping exists
\cite{ry99}. Consider the He-H interface in the SN at 2000~s, and the Cu-CH
interface in the laser experiment at 20~ns. In both, the ratio of inertial
to viscous forces (the Reynolds number) and the ratio of the convective to
conductive heat transport (the Peclet number) are large. Therefore the
viscosity and thermal diffusivity are negligible, and the dynamics of
the interface well described by Euler's equations. These equations are
invariant under a scale transformation, which maps 
lengths of  $10^{11}\rm\ cm$ into $50 \rm\ \mu m$, 
densities of $8 \times 10^{-3}\rm\ g\ cm^{-3}$ into $4 \rm\ g\ cm^{-3}$,
and pressures of $40 \rm\ Mbar$ into $0.6 \rm\ Mbar$, 
for example, at a star time of 2000~s
and a laser time of 20 nanoseconds. Thus, in a very real sense, the
experiment reconstructed a part of the supernova event.

Such experiments are also crucial for another problem: multidimensional
geometry. The experiments are inherently 3D, but can be configured to
give primarily 2D behavior. The computational load for 3D scales
from 2D  as the number of zones in the new dimension. At present, a
single workstation can easily produce 2D simulatons with good resolution,
but 3D requires parallel processing. For the next few years 
we will need to explore with 2D while we develop the capability of 
doing reliable and resolved 3D simulations with ease. The laser 
experiments can help us discover the qualitative
and quantitative limitations of 2D in real world situations.

Laboratory experiments can have other impacts on explosive nucleosynthesis.
Experiments modelling turbulent mixing, combustion, and flame
propogation are needed. However, because the scales of the systems
are so different, care must be taken in mapping the experiments into
the astrophysical domain.

\section{Thermonuclear Supernovae}

Supernovae of Type~Ia are thought to be produced in a white dwarf star
by a runaway thermonuclear reaction. They produce radioactive \nuc{56}{Ni},
and its decay to \nuc{56}{Co} and thence to \nuc{56}{Fe} give the 
characteristic light curve \cite{da96}. 
SNIa's are a major source of \nuc{56}{Fe} and other iron-group nuclei.
Because of the empirical relationship
between their brightness and the width of the luminosity peak (the Phillips
relation), they are the best distance indicators now known, and of fundamental
importance for cosmology \cite{mp93,rpk95,sp97}.

Supernovae of Type Ia present several outstanding puzzles.
\begin{itemize}
\item What are their progenitors?
\item How do they evolve to ignition?
\item How do they evolve from ignition to explosion?
\item Why does the Phillips relation between brightness and peak width work?
\end{itemize}
 
At present there is no unique and satisfactory scenario for their evolution 
up to explosion. For example, SNIa's may result from one member of
a binary pair growing in mass from accretion by its companion (there are
many possibilities for the nature of that companion and of the matter 
accreted), or by the merger of a pair of white dwarfs (which is an inherently
3D problem, and not yet computable for secular time scales and good 
resolution). 

If the accretion model is the correct one, how does it ignite? The favorite
notion is that it ignites \nuc{12}{C} at the core of the white dwarf.
To avoid
collapse or excessive production of neutron rich isotopes, the ignition
must not occur at densities which are too high 
($\rho >> 2 \times 10^9 \rm\ g/cc$). From ignition to thermal runaway there
is a period of about $10^3$ years, during which a convective region with
Urca cooling probably evolves ( see references and discussion in 
\cite{da96,mo96}),
but such a process has only been simulated in 1D.
What occurs in this ``lost millenium'' remains puzzling. See also \cite{rgi99}
for simulations of the closely related problem of the core evolution of a
11~$M_\odot$ star.

If the burning does proceed to thermal runaway, the following evolution
remains a subject of debate \cite{nh95,sew94,kh91}. It may take a new
generation of simulations to resolve the issue.

The merger scenario also has uncertainty regarding the cause of explosion.
Many discussions ( e.g., \cite{ii99}) assume that given a combined mass 
above the chandrasekhar limit,
explosion must ensue. This is untrue. Collapse or benign ignition are
also possible, maybe even more likely \cite{ml90,sn98}. 
Pioneering efforts to simulate
the merger process  probably need better resolution and longer evolutionary
times to get at this problem, or the related one of mergers involving other
combinations of constituents (white dwarfs, neutron stars, black holes) 
\cite{dbh94,scm97,fr94,mr97,rj99}.

Presumably from all this confusion will emerge a natural and convincing
reason for the Phillips relation.

\section{Almost Explosive Burning --- Setting the Stage}

Consider an evolved massive star, nearing core collapse.
Its oxygen burning shell is an important region for explosive nucleosynthesis: 
this layer is the site of explosive oxygen and silicon burning
as it is ejected by the supernova shock. Its formation and development
set the stage for the collapse of the burned core to form a neutron star
or black hole. Any discussion of core collapse, explosion mechanisms, or
continued collapse to a black hole must presume characteristics of this
formation and development, an issue we will return to below.

Almost all simulations of the stages prior to core collapse have been
one dimensional (1D), assuming spherical symmetry and instantaneous mixing
of radial layers over all angles. In 2D it is possible to
treat convection as a real hydrodynamic process, although the vortices
are pegged to the grid by assumption. Such simulations \cite{ba98,aa99}
are qualitatively different from the previous 1D ones.

This should not be surprising. The time scales are very short for such a
large object.
The duration of shell oxygen burning is only
$10^4$ seconds or so, 
the connective turnover time about $10^2$ seconds, and the sound
travel time about 1 second.
The sound speed is about 0.01 of lightspeed.
 There is little time for the convection or
the burning to settle into a steady state, or to make the zone well
mixed in composition (or even in heat), contrary to the assumptions built
into the stellar evolutionary codes. Strong downdrafts develop,
and the convection is nonlinear and nonsymmetric with regard to up and
down flows. The convective mach numbers approach tenths,
and the perturbations in pressure and density are of the order of tens
of percent at the flame zone and at the interface at the top of the formally
convective zone. The burning is sporadic and flashy. 
Perhaps the most erroneous aspect of the 1D codes is their
treatment of the boundary conditions on convection.
Material moves across
formally stable regions; in the 2D simulations \nuc{12}{C} was entrained
from across the outer interface of the oxygen convective zone, and brought
down into the flame zone where it flashed vigorously. This occurred
after about 400 seconds in the Bazan-Arnett computation \cite{ba98}.

These results have now been confirmed and extended by a completely independent
hydrodynamic code and method \cite{aa99}. At 900 seconds there seems
to be a new state developing, strongly dynamic but roughly steady on
average. The nuclear luminosity increased more than a factor of 30
above the value obtained in the 1D simulations. The burning
resembles a series of ``mini-explosions'', in which the burning happens
at higher temperatures in flashes, separated by relatively quiet intervals.

The most obvious conclusion is that the 1D simulations of
the final stages of massive stars are unrealistic, and their degree of
relevance is in question, at least regarding details of presupernovae
and pre-explosion nucleosynthesis. We can already see that the shell
luminosities are incorrect, as are the mixing algorithms. This brings
into question the pre-collapse states hitherto used for core collapse
simulations. These simulations show a dependence upon the neutronization
(that is, upon $ Y_e$), the mass of the burned core, and its entropy.   
All these features may change. 

While these new simulations do provide a first hydrodynamic description
of convection for this evolutionary stage, 
they must be improved. First, they should be pushed all the
way to core collapse, in order to determine the quantitative extent of the
changes. Second, their 2D nature may be suspect. Rotation and
magnetic fields, unavoidable is this stellar plasma,
may reduce the geometric complexity, tending back toward 2D. 
On the other hand, vortex wandering in 3D
may reduce the effects seen in 2D. This challenge is becoming
tractable with progress in computer hardware and software.

It may be that the notion, that the extent of the burning shell is determined
by the local adiabatic gradient, is flawed. The convective velocity field
is certainly NOT local. Rather, the depth to which a blob sinks depends
upon which fuels it has to flash and reverse its descent, upon
how low its entropy drops due to previous
neutrino cooling, and upon its history of electron capture. The actual
compositional structure may be better thought of as an ensemble average
of blobs being subjected to these effects. This view would imply
revisions of hydrostatic
as well and explosive nucleosynthesis yields, at least in detail, and
perhaps in general.

In this picture, an important argument against significant rotation is 
removed. The abundances in a rapidly stirred region would be representative
of the flame zone in that region, so that rotational mixing would tend to
destroy the compositional layering needed to reproduce the solar system
abundance pattern. However, if blobs were self limiting in their motion,
depending upon their composition, the layering would represent both the
temperature and compositon, and could survive.

\section{Core Collapse Supernovae}

The dramatic rotational symmetry of the rings of SN1987A, and of the eruption
of $\eta$ Carina, suggest that rotation is important for at least the late
stages of evolution of massive stars. Given this hint, let us re-examine
how rotation might and might not have consequences for supernovae. 
It appears that one of
the worst problems, the destruction of compositonal layering just discussed,
may be moot with the new view of convective burning in the presupernova.

\subsection{The Neutrino Diffusion Model}
In the Colgate model of core collapse \cite{cw66}, the collapsing core was
supposed to be both thick and thin to neutrinos. It had to be thick so that
neutrinos were copiously produced, but thin so they would stream out, but
thick enough again to deposit energy as they escaped. The neutrino transport
was not actually calculated in \cite{cw66}, but assumed to work in this 
fashion. That is, the neutrinos were assumed to diffuse quickly out of the
collapsed core, and half of their energy was deposited in the mantle.
All the models so calculated gave violent explosions; no black holes
were formed.

The first radiation hydrodynamic calculations \cite{da66,da67}
showed that if such fine tuning were allowed, explosions did result, but
also showed that black hole formation with no explosion was also likely. This
follows from a simple argument.
The diffusion time out of a homogeneous sphere of mass $M$ is
\begin{equation}
\tau_{dif} = {3 \over \pi^2c}\kappa ( 3M/4\pi)^{2/3} \rho^{1/3},
\end{equation}
where $\kappa$ is the neutrino opacity and $\rho$ the density.
The collapse becomes supersonic, so that it takes a time which is of 
the order of
and scales with the free fall time, $ \tau_{ff} \propto R/v_{ff}$.
Since $v_{ff}^2 = 2 G M / R$, we have $\tau_{c} \propto \rho^{-1/2}$.
The degree of neutrino trapping depends upon the ratio of diffusion
time to collapse time,
\begin{equation}
\tau_{dif} / \tau_{c} \propto \kappa \rho^{5/6} M^{2/3}.
\end{equation}
Increasing the neutrino opacity, the density, or the core mass $M$
tends to increase the neutrino trapping, and reduce the chance of explosion.
Massive cores tended to make black holes. With the advent of the neutral
current theory of weak interactions, the effective value of the neutrino
opacity increased, increasing trapping. Modifying the inital models or the
nuclear equation of state to give higher density at bounce also increased
trapping. If the neutrinos are trapped in the core, the collapse continues
on to black hole formation.
Nor does arbitrary tuning of $\kappa$ fix things. If the neutrino opacity
is low, the neutrinos escape but do little heating of the surrounding
and infalling mantle.

These parameters are not freely variable. The
mass $M$ is constrained by the progenitor core mass, which itself cannot
be less than the chandrasekhar mass. The weak interaction determines
$\kappa$ fairly precisely. The density of the core has varied in simulations,
but more care, realistic evaluation of nuclear equation of state, and
inclusion of 
general relativity, give more tightly constrained values. It seems fair to
say that the neutrino diffusion model does not work with realistic physics.
See \cite{mb98} for a recent review of the status of core collapse models.

\subsection{Energetics}
The unusual supernova 1998bw, and its possible identification 
\cite{ga98,iwa98,wwh98} with
the gamma-ray burst GRB980425, suggest that the supernova mechanism should
be able to provide large explosion energies ($ E_{exp} \ge 10^{52}\rm\ erg$,
or more than 10 foe). The gravitational potential energy  is 
\begin{equation}
\Omega \approx G (4\pi/3)^{1/3} M^{5/3} \rho^{1/3},
\end{equation}
in the newtonian approximation. The minimum core density is about that
of the atomic nucleus,
\begin{equation}
\rho_{nuc} = 2 \times 10^{14}\rm\ g/cc.
\end{equation}
A schwarzschild black hole has an average density of
\begin{equation}
\rho_{bh} = 3c^6/32\pi G^3M^2 = 2.85 \times 10^{16} (M_\odot/M)^2 \rm\ g/cc,
\end{equation}
but bounce densities as low as $10^{15} \rm\ g/cc$, dynamical formation
of a black hole can occur. Thus $ 2 \times 10^{14} \le \rho \le 10^{15}$
gives an estimate for the maximum energy available for explosion. In the
spherically symmetric case, black hole formation at higher masses will
limit the energy available for explosion; just below this boundary the
energy supply is at a maximum, e.g. \cite{vra79}.

\subsection{The Shock Mechanism}
If diffusion of neutrinos will not give adequate transport of energy,
shock propagation might; Bethe and Brown have led the
pursuit of this possibility
\cite{be90}. Upon reaching nuclear density, or higher, the
pressure is adequate to support the collapsing core. The size of this
region is given by equating the pressure gradient force to gravity.
The mass, for which this is true, falls as a unit, with velocity proportional
to radius, so it is called the ``homologous core.''
Prior to collapse the central density reaches 
$\rho \ge 2 \times 10^9 \rm\ g/cc$. Neutrino cooling keeps the entropy low,
so that the ``iron core mass'' approaches the chandrasekhar value,
\begin{equation}
M_{ch} / M_\odot \approx 1.45 (2 Y_e)^2,
\end{equation}
where $Y_e$ is the number of electrons per nucleon. The electron fermi
energy is several MeV. Electron capture occurs relatively slowly, but at
$Z/A = Y_e \approx 0.42$, the nuclei have a threshold for electron capture
of several Mev as well, and the neutronization is almost stopped \cite{da96}.
The smallest iron core is about $1.0 M_\odot$ for $y_e = 0.42$.
The largest homologous core would occur if no leptons were lost in subsequent
collapse, so their pressure would have the largest possible value.
At nuclear density, the difference between proton and neutron chemical
potentials will be small compared to the fermi energies of electrons and
neutrinos (of order 100 Mev), so at weak-interaction equilibrium,
\begin{equation}
\mu_e \approx \mu_\nu.
\end{equation}
This implies $ N_e/g_e = N_\nu/g_\nu$, where the helicity
phase space factors are $g_e = 2$ and $g_\nu = 1$, so $Y_e = 2 Y_\nu$,
and with no lepton escape we have for the collapsed core $Y_e = 0.28$
and $Y_\nu = 0.14$. The pressure defict is 
$ P/P_0 = (2/3)^{4/3} + {1 \over 2}(1/3)^{4/3} \approx 0.700$.
The corresponding mass deficit is $ (0.7)^{3/2} = 0.585$, so that only
about 60\% of the iron core is still in the homologous core when it
reaches nuclear density and bounces. The shock must propagate through
about 40\% of that iron core which is still infalling. For the shock
to be strong, it must dissociate the iron, requiring 8 MeV per nucleon,
or 6.4 foe ($6.4 \times 10^{51} \rm\ erg$). The energies available in such
small collapsing cores is almost always smaller than this, thus 
making the shock mechanism doubtful. 

\subsection{The Convective Mechanism}
Although the core collapse releases much more energy than seems to be necessary
for the typical supernova, the problem lies in getting it out of the nascent
neutron star/black hole. One possible solution is ``convective overturn.''
The term ``overturn'' is important because the process is unlikely to
resemble a well-developed turbulent cascade, but rather a more violent
and transient large scale turnover of the lepton-rich deep regions.
Epstein \cite{rie79} first examined the consequences of this possibility.
Early numerical simulations \cite{bbl79,lbc80} and arguments \cite{cp80}  
based on this idea were shown by Smarr et al. \cite{swbb81}  
to be overly enthusiastic, although the overturn of the outer core  would
be a generic and important feature of the core collapse.

Given the difficulties of the neutrino diffusion and the shock models,
it would appear better to allow almost all of the
iron core to fall in, then release the neutrinos at a later time.
Because there would be less mass to dissociate, this minimizes 
the dissociation losses.
The first simulation which showed such ``delayed'' behavior is due to Wilson
\cite{bw85}. This was a 1D simulation and therefore had a dubious treatment
of convective flow. It did stimulate multidimensional simulations
\cite{hbhfc94,bhf95,jm96,mcbb97} which gave results still being argued.

\subsection{The First Rotational Mechanism}
Fred Hoyle \cite{fh46,fh64} proposed that rotation played an important
role in supernovae. If there is enough angular momentum in the oxygen
shell layer which surrounds the iron core, collapse might induce explosion
by oxygen burning. This did not result in the Colgate-White simulations
because their collapse generated an excessively strong rarefaction wave
which swallowed the oxygen shell. A more careful treatment of the onset
of core collapse \cite{da77} showed a longer initial contraction, in which
the oxygen shell did burn violently (but not quite explosively). 
Following Fowler
and Hoyle's suggestion \cite{fh64}, Bodenheimer and Woosley \cite{bw83}
simulated some simple explosions of this type, with rough estimates of
the rotational state of the presupernova. These should be re-examined
using more realistic multidimensional precollapse models and better
resolution. The combined
effects of rotation and hydrodynamics on the core mass, the neutron excess,
and the entropy would be interesting in its own right. Even in the
1D case the explosion of SN1987A would have gotten about 10\% of its 
energy from explosive burning of oxygen.

\subsection{Rotation Revisited}
Rotation could cause significant effects in the core itself
\cite{mm89}.
Extreme rotation would cause the collapse to halt due to centrifugal
forces at a density less than nuclear density. While this would
aid neutrino escape, it would also release less gravitational energy
and lower the neutrino energies, making them less able to deposit energy
in the outer layers. 
An unresolved issue is the rate of angular momentum transport, which in
this case would determine the secular evolution to the neutron star or
black hole state.

Magnetic fields might be important as well \cite{fh46,go71,lbw70}. The plasma
is matter, not field, dominated (a high $\beta$ plasma), so that magnetic
fields would be subtle, at least initially. With dynamo action the field
would be strengthened, and buoyancy would tend to move it to regions
in which its effects might be still more important. This is justifiably an
old problem because it is inherently 3D and time dependent. The failure
of the pulsar models to predict a luminosity below the observed radioactive
decay of \nuc{56}{Co}
in SN1987A  may indicate that magnetic effects are not a dominant
feature; however evidence for a pulsar (or alternatively a black hole)
would clarify this point.

A more modest and perhaps realistic (?) role for rotation is to induce mixing,
but not centrifugal braking. The rotation and magnetic field would guide
the overturn, emphasizing large scale motion (low modes). This may help the
``convection'' model discussed above. 

In any case, rotation is likely to have an important effect in that it
breaks the spherical symmetry in a characteristic way.
While neutrino diffusion of energy will tend to be spherically symmetric,
rotation tends to give the rotation axis a special role. Centrifugal force
tends to evacuate these regions, so that they would have a lower density
and, if heat transport is effective, a higher entropy. Such 
polar hot spots might be conducive to the formation of jets, and trigger
overturn.
Preliminary attempts to examine the consequences \cite{kh99,mw98}
are promising.

\section{Summary}

\begin{itemize}
\item Explosive nucleosynthesis is a combination of nuclear reactions
with hydrodynamics, and depends upon explosions mechanisms for supernovae.

\item Incomplete mixing gets worse for massive stars and explosive conditions.

\item Nuclear burning, and yield predictions, is complicated by 
hydrodynamic convection in presupernovae.

\item Laboratory experiments with high intensity lasers has become a good
testing ground for astrophysics codes. With careful scaling, such experiments
can reproduce supernova phenomena.

\item Understanding of SNIa's is impeded by a lack of progenitor information,
and by theoretical problems with approach to ignition, runaway, and 
stellar merger.

\item Simulations of oxygen shell burning using actual (2D) hydrodynamics
differ drastically from 1D results. The first simulations have been confirmed
by an independent code, and is being carried further.
 
\item For core collapse explosion mechanisms, 
neutrino diffusion and prompt shock models are dead, and pure convection
models may be sick.

\item Rotation must be included in progenitor and core collapse evolution.

\item SN1998bw shows that the energy problem with core collapse supernovae
is worse than previously supposed. There are events having energies much
larger than several foe ($10^{51}\rm\ erg$).

\item Because newtonian gravity and centrifugal force are scale free,
jet formation is likely to occur on scales having effective heat flow.
This may connect  protostar jets and galactic jets (with heat flow by
radiative diffusion and convection), 
and core collapse supernovae (with heat flow by
neutrino diffusion and convection). 

\end{itemize}

The prospects are simply wonderful. Our tools are getting much better, and
may finally be up to the task of simulation of explosive
nucleosynthesis events in realistic geometry.
Computer technology, nuclear reaction rates, and hydrodynamics rates
are improving and being verified in new ways. Meteoritic data \cite{ez98}
has presented quantitative challenges to 1D model predictions. With the
impending crash of SN1987A into its rings, the renewed activity of
$\eta$ Carina, the active supernova searches out to large redshift, 
and the possibility of a connection between core collapse supernovae
and GRB's, we may expect to learn many new things.

\section*{Acknowledgment}   
This research is supported by DOE grant DE-FG03-98DP00214/A001.

\medskip

\end{document}